\documentclass[10pt,preprint]{aastex}
\usepackage{natbib}
\usepackage{amsmath}
\usepackage{color, soul}

\begin{document}
\title{The Stellar Halo and Tidal Streams of Messier~63}
\author{Shawn M. Staudaher\altaffilmark{1}, Daniel A. Dale\altaffilmark{1}, Liese van Zee\altaffilmark{2}, Kate L. Barnes\altaffilmark{2}, David O. Cook\altaffilmark{2}}
\altaffiltext{1}{Astronomy Department, University of Wyoming, WY 82071}
\altaffiltext{2}{Astronomy Department, Indiana University, IN 47405}
\email{sstaudah@uwyo.edu}

\def\HII{\ion{H}{2}}
\def\NII{[\ion{N}{2}]}
\def\ha{H$\alpha$}
\def\um{${\rm \mu}$m}
\def\spit{$\it{Spitzer}$}
\def\mips{24~$\mu$m}

\def\StreamSigma{$42 \pm 10 \arcsec$}
\def\StreamPeak{$2.0 \pm 0.3$~kJy~sr$^{-1}$}
\def\StreamWidth{$75 \pm 18 \arcsec$}
\def\StreamWidthPC{$2.9 \pm 0.7$~kpc}

\def\SBulgeM{$8 \pm 3 \times 10^{9}~M_{\odot}$}
\def\SDiskMOne{$3 \pm 1 \times 10^{10}~M_{\odot}$}
\def\SDiskMTwo{$3 \pm 1 \times 10^{10}~M_{\odot}$}
\def\SMTotal{$6 \pm 2 \times 10^{10}~M_{\odot}$}

\def\HBulgeM{$1.2 \pm 0.5 \times 10^{10}~M_{\odot}$}
\def\HDiskM{$4 \pm 2 \times 10^{10}~M_{\odot}$}
\def\HHaloM{$8 \pm 3 \times 10^{9}~M_{\odot}$}
\def\HMTotal{$6 \pm 2 \times 10^{10}~M_{\odot}$}

\def\Mfrac{$12 \pm 2 \%$}

\def\Mpoly{$1.2 \pm 0.5 \times 10^8 M_{\odot}$}
\def\Mlight{(5$\pm 2) \eta \times 10^8 M_{\odot}$}
\def\Mdyn{2.3$\pm 1.6 \times 10^8 M_{\odot}$}
\def\tdisrup{(2.3$\pm$0.7)$\eta$~Gyr}

\def\HaloStreamFrac{$16 \pm 2$}
\def\RateFrac{$3 \pm 1$}

\begin{abstract}
	
We present new near-infrared (NIR) observations of M63 from the Extended Disk Galaxy Exploration Science (EDGES) Survey.  The extremely deep 3.6~$\mu$m mosaic reaches 29~AB~mag~arcsec$^{-2}$ at the outer reaches of the azimuthally-averaged surface brightness profile.  At this depth the consequences of galactic accretion are found within a nearby tidal stream and an up-bending break in the slope of the surface brightness profile.  This break occurs at a semi-major axis length of $\sim$8\arcmin, and is evidence of either an enhanced outer disc or an inner stellar halo.  Simulations of galaxy evolution, along with our observations, support an inner halo as the explanation for the up-bending break.  The mass of this halo component is the largest found in an individual galaxy thus far.  Additionally, our observations detect a nearby tidal stream.  The mass of the stream suggests that a handful of such accretion events are necessary to populate the inner stellar halo.  We also find that the accretion rate of the galaxy from the stream alone underestimates the accretion rate required to build M63's inner stellar halo.	

\end{abstract}


\section{Introduction}

A fundamental question in galaxy evolution is: what are the consequences of a $\Lambda$CDM cosmology on today's galaxies?  Simulations based upon a $\Lambda$CDM cosmology find that major mergers, where large galaxies of roughly equivalent mass merge, were common in the early Universe.  These dramatic encounters destroy the underlying structures of the galaxies undergoing the merger, and cause an entirely new type of galaxy to form \citep{Hopkins+06, Chilingarian+10, Borlaff+14, Sonnenfeld+14}.  However, while major mergers may have governed galaxy evolution for massive galaxies, these mergers are rare in the nearby Universe.  The driving force for the evolution of today's intermediate mass galaxies is from minor mergers, where satellite galaxies are accreted on to host galaxies without destroying the underlying structure of the host.  Simulations of minor mergers between a massive spiral and a dwarf satellite find observable evidence for the merger: long-lived tidal streams (a dwarf in the process of being accreted), and a stellar halo \citep[the remnants of multiple mergers that form a halo surrounding the disc;][]{Bullock+Johnston05, Purcell+07, Cooper+10, Purcell+11, Cooper+13}.

The search for the signs of minor mergers, streams and haloes began in the Milky Way.  The earliest discovered stream was the gaseous stream from the Large and Small Magellanic clouds \citep{Wannier+72, Mathewson+74}.  Subsequently, stellar streams have been detected coming from the Sagittarius \citep{Ibata+94} dwarf galaxy.  This led to the discovery of a multitude of streams surrounding the Milky Way \citep[e.g.][]{Belokurov+06, Martin+14}. In the Milky Way's stellar halo, the largest globular clusters, such as $\omega$~Centauri, are thought to be the cores of accreted satellites \citep[e.g.][]{Johnson+10, Marino+10}.  These large clusters host multiple stellar populations, which cannot be formed in a simple Milky Way cluster \citep{Mackey+10, Milone+10, Gratton+12}.

Despite the success with the Milky Way, the detection of extragalactic stellar streams and haloes has proven difficult with ground-based intruments.  The Milky Way studies have relied on detecting very faint individual stars, a task that is next to impossible for extragalactic sources from the ground without an 8~m class telescope.  Thus, most ground-based extragalactic studies beyond the Local Group are limited to studying the surface brightness of streams and haloes.  This remains a difficult task; simulations find that the surface brightnesses required to detect streams and haloes are below 28~AB~mag~arcsec$^{-2}$ \citep{Purcell+07, Cooper+10, Cooper+13}.  Reaching these extreme depths is complicated by the limited amounts of time available on research-grade telescopes, difficulties in obtaining accurate flat fields, and the extended wings of the point spread functions of reflecting telescopes.  The works of \cite{Martinez-Delgado+08, Martinez-Delgado+09, Martinez-Delgado+10, Martinez-Delgado+14} used a small telescope with a clear filter (essentially Sloan $g$, $r$, and $i$ combined) to detect the tidal features of nearby galaxies.  Subsequently, these detections justified follow-up observations with larger telescopes \citep{Chonis+11, Martinez-Delgado+12}; these studies derived the physical properties of the tidal streams such as their masses and star formation histories.  Many of the above issues with low surface brightness studies are solved by using an array of refracting telescopes.  The robotic Dragonfly array \citep{Abraham+vanDokkum14} is currently surveying, to very low surface brightness, a large number of nearby galaxies.  Interestingly, \cite{Abraham+vanDokkum14} found only a hint of a stellar halo around M101, down to $\mu_g\sim32$~AB~mag~arcsec$^{-2}$.  Finally, by deprojecting, and then stacking, many thousands of SDSS galaxies, \cite{D'Souza+14} have detected the average stellar halo of spiral galaxies with masses from $10^{10}-10^{11}~M_{\odot}$.

Space-based telescopes may be the best way to detect stellar haloes and streams. The Galaxy Halos, Outer disks, Substructure, Thick disks and Star clusters (GHOSTS) survey \citep{Radburn-Smith+11} is investigating the haloes of several nearby spirals with the Hubble Space Telescope.  These data have successfully detected the stellar halo of NGC~253 \citep{Bailin+11}, and a mass for the stellar halo has been derived.  With the cryogen of the Spitzer Space Telescope exhausted, the telescope has entered a warm mission with only the 3.6~\um\ and 4.5~\um\ cameras active.  These bands are excellent for low stellar mass surface density studies of older stellar populations down to $\sim$0.2~$M_{\odot}~pc^{-2}$ \citep{Krick+11}.  This is due to the insensitivity of the 3.6 and 4.5~\um\ $M/L$ ratio of the older stellar populations to the effects of extinction, metallicity, and stellar evolution \citep{Bell+deJong01, Oh+08, Meidt+14}.  The majority of the uncertainty in the 3.6~\um\ $M/L$ ratio comes from contaminants such as emission from polyaromatic hydrocarbons and short-lived asymptotic giant branch stars.  Several investigations \citep{vanZee+09, Courteau+11, Krick+11, vanZee+11, Barnes+14, Bush+14} have taken advantage of the warm mission to conduct ultra-deep low surface brightness studies. In particular, \cite{Courteau+11} have measured the mass of the stellar halo in the Andromeda galaxy, and \cite{Barnes+14} have measured the physical properties of a tidal stream near M83.

In this work we analyse the near-infrared (NIR) low surface brightness properties of M63 (NGC5055), a member of the Extended Disk Galaxy Exploration Science (EDGES) Survey \citep{vanZee+11}.  Our 3.6~\um\ and 4.5~\um\ mosaics showcase the results of a minor merger between M63 and a nearby satellite.  There is a clearly visible tidal stream and stellar halo associated with this galaxy.  We have measured the properties of the stream and stellar halo that we compare to models based on hierarchical galaxy evolution in a $\Lambda$CDM universe.

\section{Near-Infrared Imaging}
\label{sec:nir}
The hallmark of the EDGES survey is ultra-deep (29 AB~mag~arcsec$^{-2}$) wide-field ($>$5 $R_{25}$) Spitzer near-infrared (NIR) mosaics of 92 nearby galaxies.  These data were taken during Spitzer's Cycle 8, a warm mission Exploration Science program (van Zee et al. 2011; PID 80025).  In the following subsections we detail the steps taken to construct the 3.6~\um\ and 4.5~\um\ mosaics for M63.

\subsection{Maps}
\label{sec:maps}
The extent of the EDGES M63 mosaic, ($>$5 $R_{25}$), is 31\farcm5 \citep{deVaucouleurs+91}; significantly larger than the extent of the IRAC field of view, $5\farcm2\times5\farcm2$.  Therefore, the observing program for M63 was designed to create mosaics from a grid-like mapping pattern of many individual dithered pointings.  This mapping pattern was set up to have the 3.6~\um\ mosaic centred on the target, with the 4.5~\um\ data being taken concurrently.  The 4.5~\um\ channel is separated from the 3.6~\um\ channel by 6$\farcm$5 and thus the 4.5~\um\ map is slightly offset from the 3.6~\um\ map.  The M63 mosaic is composed of 2,568 individual exposures to reach our target depth of 1800~s~pixel$^{-1}$ while minimizing effects of saturation.  The images were taken in four Astronomical Observation Requests (AORs), sets of 400 to 800 exposures on March 29, 2012 and August 25, 2012 (Keys: 44244992, 44245248, 44246016, and 44246272).  During each date two AORs were taken in quick succession, with a break to begin a new AOR.  This strategy assists in asteroid and artefact removal, and also extends the 4.5~\um\ coverage by allowing the 4.5~\um\ images to be taken at different telescope roll angles.

\subsection{Data Processing}
\label{sec:data_proc}
Before mosaicking, a number of pre-processing operations are applied to the individual frames to remove artefacts common to warm mission data.  The first correction removes slew residuals caused by tracking over bright sources without a shutter.  These slew residuals create sub-structure in every frame of an AOR.  This structure is removed by subtracting a median-frame (constructed by stacking every frame in the AOR into a 3D data cube and taking a median down the frame number axis) from every image in the AOR.

Bright sources also create a complex negative bias in the bright source's columns, the so-called ``column pulldown effect''.  In the cryogenic mission this effect creates a single-valued bias over an entire column which may be reliably removed.  However, in the warm mission this effect is more difficult to remove as the bias value depends on distance from the centre of the source.  The Spitzer Science Center and various third parties have published algorithms which correct for the warm mission column-pulldown, but they are not optimized for the EDGES dataset, which contains diffuse structure and longer exposure times than average.  We have developed a pulldown corrector specifically for EDGES.  Our pulldown corrector first identifies pulled-down columns by flagging columns with bright sources.  Then, it estimates the true value of the column as the average of the flanking  columns' rows.  This estimated column is then subtracted from the actual column.  The result is the functional form of the column pulldown.  This result is fit with a linear function and the linear function is subtracted from the actual column, which corrects the pulldown.  By using our pulldown corrector along with the Spitzer Science Center's corrector, the majority of column pulldown artefacts are removed automatically.  The remaining pulled-down columns are manually flagged and removed.  The damage done to the mosaic by their removal is minimized by the excellent coverage of EDGES, where most pixels contain data from 18 separate pointings.

\begin{figure}
	\epsscale{0.8}
	\plotone{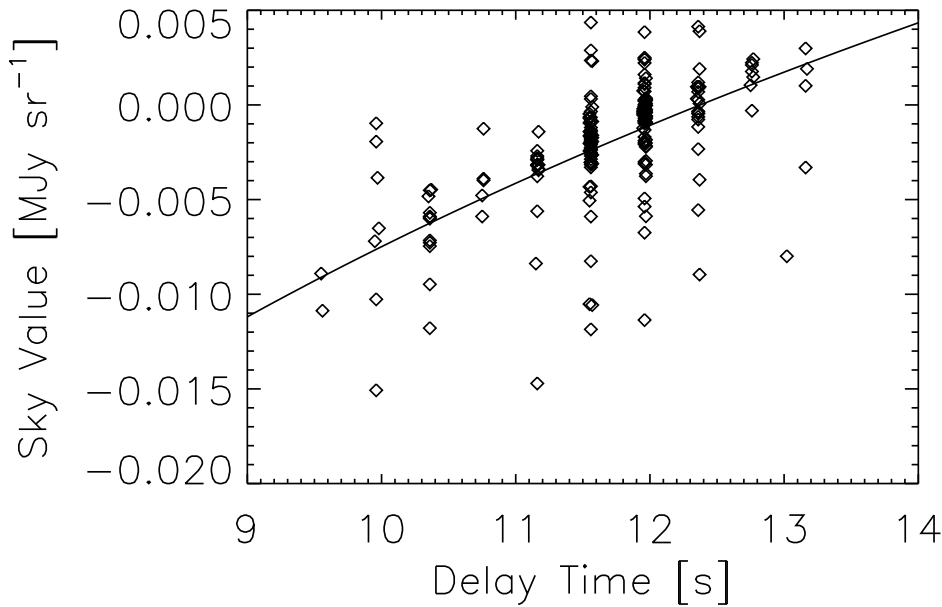}
	\caption{The first frame effect due to delay time for AOR key 44244992, the first AOR taken for M63.  The solid line is the exponential fit used to correct the effect as discussed in \S\ref{sec:data_proc}.}
	\label{fig:first_frame}
\end{figure}

In addition to the column bias issues, there are two sources of frame-wide bias.  As discussed in \cite{Krick+11}, the overall bias level of the individual frames depends on the delay time between exposures, the so-called ``first frame effect''.  The AORs for M63 have enough frames to measure this effect for each of the four AORs individually.  This is accomplished by measuring the background level of every frame with the IDL task {\tt SKY} and fitting a logarithmic profile to the background as a function of delay time with the IDL task {\tt MPFITFUN} \citep{Markwardt09} (see Figure~\ref{fig:first_frame}) of the form:
\begin{equation}
	c_{\rm ff}(t_{\rm delay})=c_0{\rm log}(t_{\rm delay})+c_1,
\end{equation}
where $c_{\rm ff}$ is the bias level due to the ``first frame effect'', $t_{\rm delay}$ is the delay time between exposures, and $c_{0,1}$ are the free parameters in the fit. This correction is applied to every frame in an individual AOR.

\begin{figure}
	\epsscale{0.8}
	\plotone{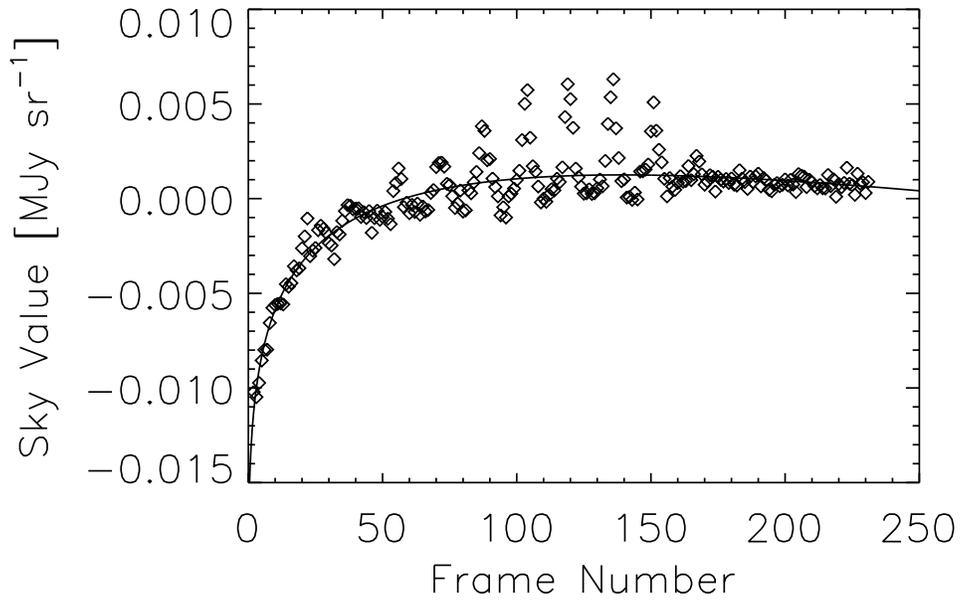}
	\caption{The buildup effect due to frame number for AOR key 44244992, the first AOR taken for M63.  The solid line is the polynomial fit used to correct the effect as discussed in \S\ref{sec:data_proc}. Note that the scatter seen in frames 100-170 corresponds to frames in which M63 fills the field-of-view, so the measured sky value includes both sky and galaxy light.  By fitting a functional form to the sky value, rather than using the local sky for each frame, we retain the extended, diffuse light associated with this galaxy.}
	\label{fig:buildup}
\end{figure}

\cite{Krick+11} also found that the bias level depends on frame number; as an AOR progresses the bias level of individual frames increases.  This effect is greatest for the first few frames, and again the M63 AORs are large enough to measure this effect for each AOR (see Figure~\ref{fig:buildup}).  Following \cite{Krick+11}, this effect is measured by fitting roots to the background as a function of frame number with the IDL task {\tt MPFITFUN}, of the form:
\begin{equation}
	c_{\rm b}(n_{\rm frame})=c_0n_{\rm frame}^{1/4}+c_1n_{\rm frame}^{1/3}+c_2n_{\rm frame}^{1/2},
\end{equation}
where $c_{\rm b}$ is the bias level due to frame buildup, $n_{\rm frame}$ is the frame number, and $c_0$, $c_1$, and $c_2$ are the free parameters in the fit.  This correction is applied to every frame in an individual AOR.

Once these artefacts have been removed, the AORs are processed individually with the standard MOPEX pipeline.  The plate scale is set to 0.75~arcsec~pixel$^{-1}$, and the AORs are set to a common astrometrical solution.  The diffuse background of each AOR is fit as a plane gradient to manually defined regions.  This plane gradient is subtracted from each AOR, and because of the common astrometrical solution, the AORs are simply averaged to produce the final mosaic.

Due to the lack of short exposures, the central portions of the M63 mosaics suffer from saturated pixels.  To remedy this, archival data from the Spitzer Infrared Nearby Galaxies Survey are used to create a mosaic with the same astrometric solution as the EDGES mosaic.  The saturated pixels in the EDGES mosaic are replaced with the archival data after matching background levels.

\section{Analysis}
\label{sec:analysis}
M63 (see Figure~\ref{fig:mosaic}) is a flocculent spiral at 7.9~Mpc \citep{Tully+09, Nasonova+11}.  It features an upbending radial profile \citep[][see also Figure~\ref{fig:surf}]{Chonis+11}, which may be a stellar halo \citep{Purcell+07, Purcell+11, Cooper+13}.  Additionally, there is evidence of an ongoing accretion event with the presence of a tidal stream \citep{vanderKruit79, Chonis+11}.  These features make M63 an excellent test bed for the power of the EDGES survey.  We measure the surface brightness profile of M63 and convert that to a mass density profile.  We also fit S\'ersic functions to the data and measure the masses of the individual components of M63, the bulge and two discs.  In addition, we fit a S\'ersic disc and bulge along with a power-law halo component.  The nearby tidal stream is present in the 3.6~\um\ mosaic, and we measure the physical properties of the stream, including the stellar mass of the progenitor galaxy, and the time since the progenitor was disrupted.

\subsection{Surface Brightness Profiles}
\label{sec:surf_bright_prof}
\begin{figure}
	\epsscale{0.7}
	\plotone{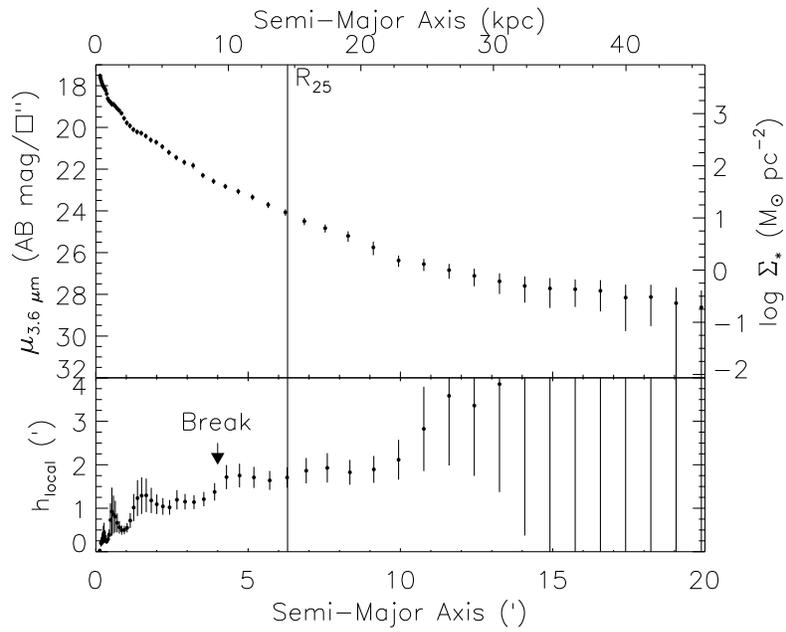}
	\caption{The upper plot is the surface brightness profile of M63, see \S\ref{sec:surf_bright_prof} for details. The x-axis is in arcminutes and also in kpc.  The y-axis is in terms of a surface brightness in mag~arcsec$^{-2}$ and a mass surface density in M$_{\odot} {\rm pc^{-2}}$. The vertical line is the RC3 value of $R_{25}$ \citep{deVaucouleurs+91}.  The lower plot is the local scale length versus radius (See \S\ref{sec:prof_fit}); the arrow designates where a possible up-bending break occurs.}
	\label{fig:surf}
\end{figure}

We present the 3.6~\um\ surface brightness profile for M63 in Figure~\ref{fig:surf}.  Before the surface brightness profile was measured, the mosaic was prepared by removing point sources and smoothing.  The point sources were removed in an iterative process, starting with a H{\" o}gbom CLEAN algorithm, then masking with SExtractor, and finally removing any remaining sources by hand.  The smoothing was accomplished with a Gaussian filter with a standard deviation of 6~pixels, with the IDL function {\tt GAUSS\_SMOOTH}.

After the mosaic was prepared, the surface brightness profile was measured by fitting elliptical isophotes to the data using the IRAF task {\tt ELLIPSE} for the inner 9$\arcmin$.  Beyond 9$\arcmin$, {\tt ELLIPSE} fails to find a solution due to low signal to noise.  The surface brightness profile beyond 9$\arcmin$ is measured with fixed elliptical annuli defined by the final isophote from {\tt ELLIPSE}.  This isophote has an axial ratio of 0.62, position angle of $103$\degr\ east of north, with a major axis of 9\arcmin.

Our surface brightness profiles are aperture corrected with the calibration from the Spitzer Science Center's IRAC instrument handbook \cite[see][]{Dale+09}.  This correction takes the form:
\begin{equation}
	\frac{f_{\rm true}}{f_{\rm measured}}=A \times {\rm exp}(-R^B)+C,
\end{equation}
where $f$ is the flux density measured within an aperture, $A$, $B$, and $C$ are constants, and $R$ is the radius of the aperture.  To use this correction with the annuli of the surface brightness profile, we measure the annuli as the difference between two apertures with a difference in semi-major axis of one pixel.

To convert the surface brightness profile from MJy~sr$^{-1}$, calibrated from the MOPEX pipeline, into a Vega-magnitude based surface brightness, we use the calibration from \cite{Reach+05} ($c_m^{3.6}=$280.9$\pm$4.1~Jy).  Furthermore, we calculate a mass surface density via:
\begin{equation}
	\Sigma[M_{\odot} {\rm pc^{-2}}] = \Upsilon_{\star}^{3.6} 10^{-0.4(\mu_{3.6} - M_{\odot}^{3.6} - 21.572)},
\end{equation}
where $\Sigma$ is the mass surface density, $\Upsilon_{\star}^{3.6}=0.5$ is the stellar mass-to-light ratio at 3.6~\um\ \citep{Oh+08, Eskew+12, Barnes+14, Meidt+14}, $\mu_{3.6}$ is the measured surface brightness in Vega~mag~arcsec$^{-2}$, and $M_{\odot}^{3.6}=3.24$ is the absolute magnitude of the Sun at 3.6~\um\ \citep{Oh+08}, and 21.572 is the factor to convert from per square arcseconds to per square parsecs.  See Figure~\ref{fig:surf} for the result.  Note that the uncertainties are the RMS value reported by {\tt ELLIPSE}, or the standard deviation within the annulus for data beyond 9\arcmin.

\subsection{Profile Fitting and Mass Estimates of Fit Components}
\label{sec:prof_fit}
\begin{figure}
	\epsscale{0.7}
	\plotone{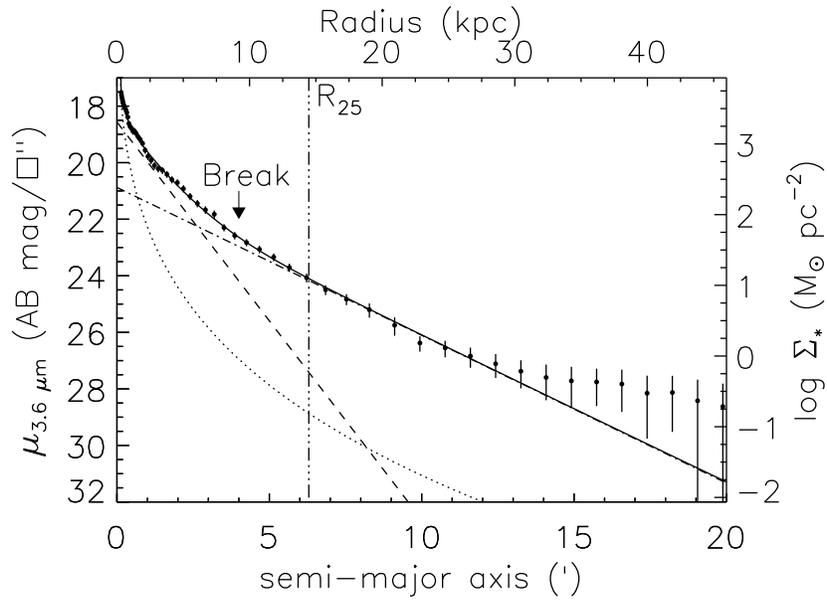}
	\caption{The surface brightness profile of M63 fit with a S\'ersic bulge and two discs as discussed in \S\ref{sec:prof_fit}.  The solid line is the total fit.  The dotted line is the fit to the bulge component.  The dashed line is the fit to the inner disc.  The dashed-dotted line is the fit to the outer disc.  Table~\ref{table:fit} summarizes the parameters of these fits.  The vertical line is the RC3 value of $R_{25}$ \citep{deVaucouleurs+91} \and the arrow designates approximately where the up-bending break occurs.}
	\label{fig:disk_fit}
\end{figure}

We have fit two models to our surface brightness profiles; the first is a S\'ersic bulge with two S\'ersic discs, and the second is a S\'ersic bulge and disc with a halo component described by a power law.  For the first model we use the following equation \citep{Sersic63, Sersic68} for each individual component:
\begin{equation}
	S(R)=S_0{\rm exp}[-(R/h)^{1/n}],
\end{equation}
where $S$ is surface brightness in MJy~sr$^{-1}$, $R$ is the radius in arcmin, $h$ is the scale length in arcmin, and $n$ is the S\'ersic index.  We use standard S\'ersic indices: $n=4$ for the bulge, and $n=1$ for the two disc components.  The IDL Levenburg-Marquardt fitting algorithm {\tt MPFITFUN} \citep{Markwardt09} fits the model to the data.  Table~\ref{table:fit} summarizes the results of the profile fit, and Figure~\ref{fig:disk_fit} shows these fits.  Additionally, to illustrate that the profile requires two discdisc components, Figure~\ref{fig:surf} contains a local scale length profile of M63.  These local scale lengths are computed by fitting a single disc component to five consecutive points on the surface brightness profile.  From the local scale length profile, and the fit, M63 at 3.6~\um\ is best described by a two disc up-bending profile, where the slope increases at the bend radius, with the bend at $\sim$4\farcm5.

\begin{figure}
	\epsscale{0.7}
	\plotone{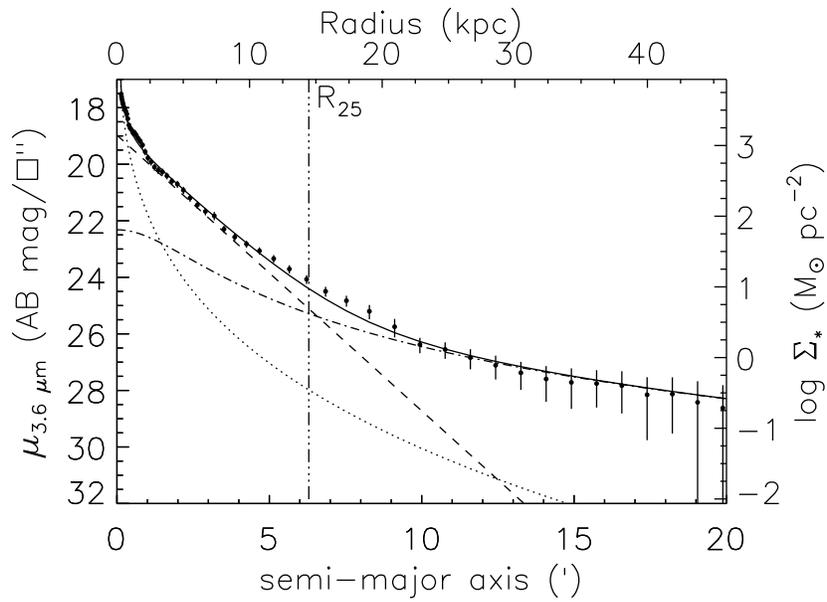}
	\caption{The surface brightness profile of M63 fit with a S\'ersic bulge and disc along with a fit to a stellar halo as a power law from \cite{Courteau+11}.  The solid line is the total fit.  The dotted line is the fit to the bulge.  The dashed line is the fit to the disc.  The dashed-dotted line is the fit to the halo. Table~\ref{table:fit} summarizes the parameters of the fit.  The vertical line is the RC3 value of $R_{25}$ \citep{deVaucouleurs+91}.}
	\label{fig:halo_fit}
\end{figure}

Up-bending breaks at large radii may also result from the detection of an inner stellar halo.  For this scenario we fit our surface brightness profile with a S\'ersic bulge and disc, and the {\it U}-model from \cite{Courteau+11}, a power law designed to measure the inner stellar halo:
\begin{equation}
	S_h(R)=S_*\left\{\frac{1+(R_*/a_h)^2}{1+(R/a_h)^2}\right\}^{\alpha},
\end{equation}
where $S_h$ is the surface brightness of the inner halo component in MJy~sr$^{-1}$ at a radius $R$, $R_*$ is the turnover radius (30~kpc), $\alpha=1.26\pm0.4$, $a_h=5.2\pm0.16$~kpc and $S_h(R_*)=S_*$.  We fit this model as above. The results of this fit are found in Table~\ref{table:fit} and Figure~\ref{fig:halo_fit}.

From our model fits we measure the mass of the individual components of M63.  For the S\'ersic components this is accomplished by first integrating over $R$ to find the total flux density, from \cite{Graham+Driver05}:
\begin{equation}
	f(<R)=2 \pi \frac{b}{a} S_0 h^2 n \gamma(2n,x),
\end{equation}
where $f(<R)$ is the total flux density up to a radius $R$, $a$ and $b$ are the semi-major and semi-minor axes respectively, $\gamma$ is the incomplete gamma function, and $x=(R/h)^{1/n}$. We use the average $b/a$ found by our ellipse fitting, and $R=200$~kpc.  To find the flux density of the power law fit we use the following from \cite{Courteau+11}:
\begin{equation}
	f_h(<R) = 2 \pi \frac{b}{a} S_*  R_*^2 \frac{1+a_*^2}{2(\alpha-1)} \times \left\{\left(\frac{1+a_*^2}{a_*^2}\right)^{\alpha-1} - \left(\frac{1+a_*^2}{s_{max}^2+a_*^2}\right)^{\alpha-1}\right\},
\end{equation}
where $a_*=a_h/R_*$ and $s_{max}=R/R_*$.  To convert the flux densities to a mass we use:
\begin{equation}
	M_{\rm total}[M_{\odot}]= \Upsilon_{\star}^{3.6}\frac{f{\rm[Jy]}}{c_m^{3.6}\rm[Jy]} \times 10^{0.4(\mu + M_{\odot}^{3.6})},
	\label{eq:mass_conversion}
\end{equation}
where $M_{\rm total}$ is the total mass, $\Upsilon_{\star}^{3.6}=0.5$ is the stellar mass-to-light ratio at 3.6~\um\ \citep{Oh+08, Eskew+12, Barnes+14, Meidt+14}, $c_m^{3.6}=280.9$~Jy is the 3.6~\um\ calibration factor to Vega-magnitudes \citep{Reach+05}, $\mu=29.49 \pm 0.35$~mag is the distance modulus \citep{Tully+09, Nasonova+11}, and $M_{\odot}^{3.6}=3.24$~mag is the absolute magnitude of the Sun at 3.6~\um\ \citep{Oh+08}.  See Table~\ref{table:fit} for the results.

\subsection{Tidal Stream}
\label{sec:tidal_stream}

\begin{figure}
	\epsscale{1.1}
	\plottwo{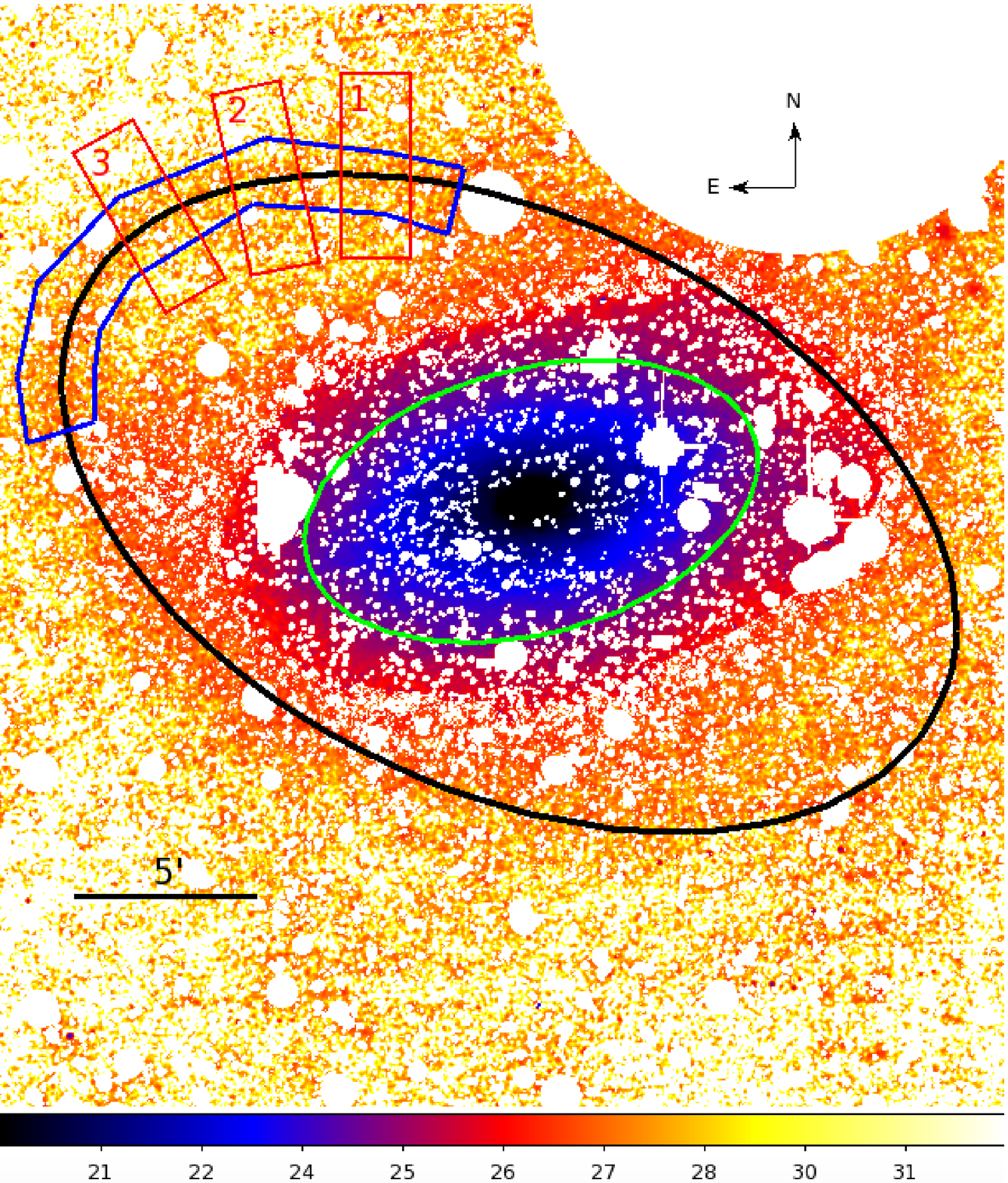}{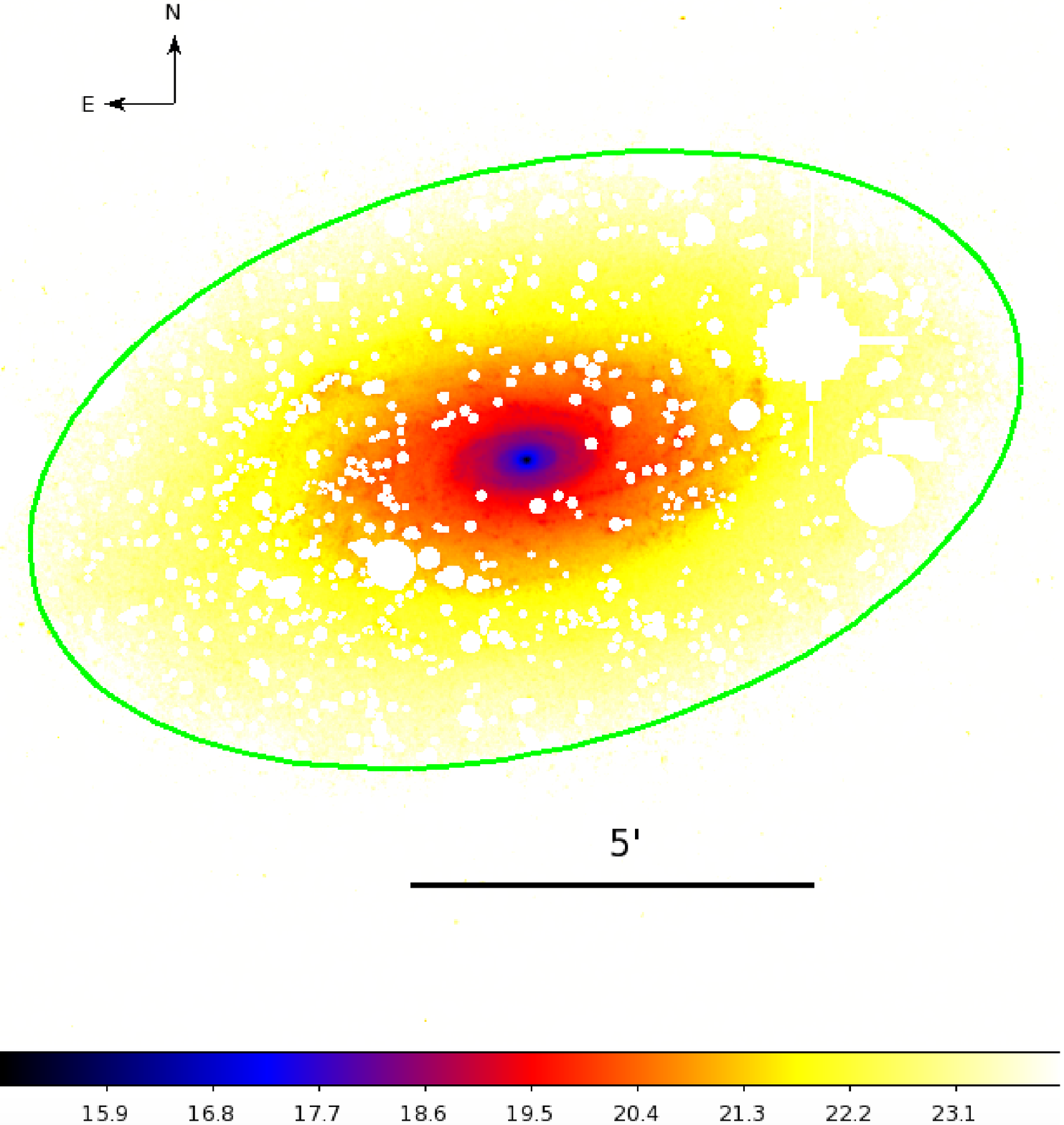}
	\caption{The 3.6~\um\ mosaic of M63; calibrated in AB~mag~arcsec$^{-2}$ according to the coloured bar.  The image on the left shows the entire extent of the galaxy while the image on the right is a zoomed-in view which highlights the traditional optical extent of the galaxy.  White pixels are masked from the analysis.  The green ellipse is the RC3 $R_{25}$ ellipse \citep{deVaucouleurs+91}.  The black ellipse is our fit to the stellar stream.  The blue box is the polygonal aperture used in \S\ref{sec:tidal_stream} to measure the mass of the stream.  The red rectangles are used to measure the width of the stream, as discussed in \S\ref{sec:tidal_stream}.}
	\label{fig:mosaic}
\end{figure}

\begin{figure}
	\epsscale{0.7}
	\plotone{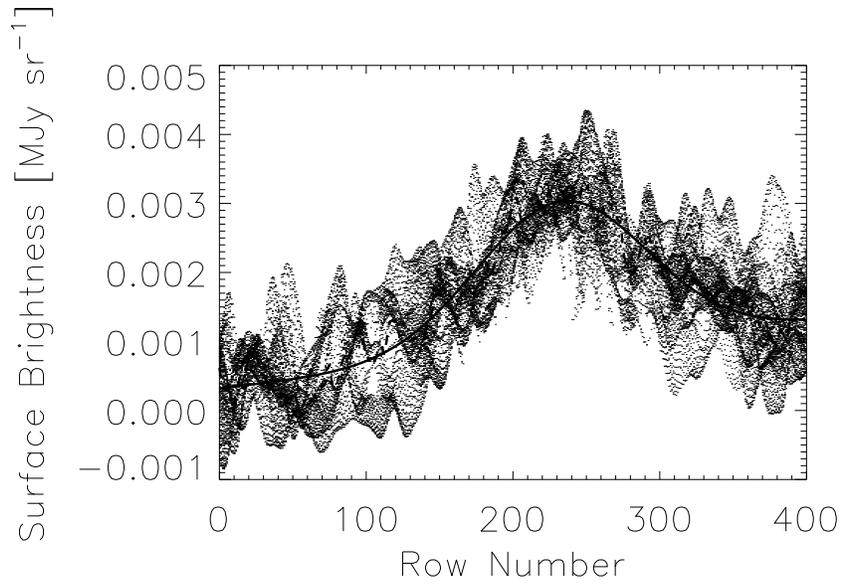}
	\caption{Surface brightness versus row number for the second rectangle on M63's tidal stream (see Figure~\ref{fig:mosaic}).  The points are individual pixels bounded within the rectangle.  The dashed line is the median of the pixels along a row.  The solid line is the Gaussian and linear fit to the median.  We use the standard deviation from the Gaussian fit as an analog to the width of the stream.  See \S\ref{sec:tidal_stream} for more information.}
	\label{fig:box}
\end{figure}

M63 hosts a nearby tidal stream, first discovered by \cite{vanderKruit79}, and recently studied in-depth by \cite{Chonis+11}.  The portion of the stream unobscured by the light from M63's disc and halo is fit well by an ellipse, as seen in Figure~\ref{fig:mosaic}.  This ellipse has a position angle of 75\degr\ east of north, an axial ratio of $b/a=0.64$, and a semi-major axis of $a=13\farcm5$ (31~kpc at our fiducial distance to M63 of 7.9~Mpc).  To measure the width of the stream, we have defined three rectangular apertures over three sections of the stream, seen in Figure~\ref{fig:mosaic}.  To produce a 2-D profile of the stream, the rows perpendicular to north are averaged, see Figure~\ref{fig:box}.  A three-parameter Gaussian (peak, position, and standard deviation) and a linear function is then fit to the 2-D profile.  From the average of the three apertures, the stream has a peak of \StreamPeak, with a standard deviation of \StreamSigma.  Following the example of \cite{Barnes+14}, where the width of the stream is 20\% of the maximum, results in a width of \StreamWidth, or \StreamWidthPC.

To measure the total mass of the stream we employ three methods.  For the first, we simply measure the total surface brightness within a polygonal aperture,  and convert the result to a mass using Equation~\ref{eq:mass_conversion}.  To account for the missing light from masked stars and galaxies, the result is multipled by the ratio of total area to missing area (1.29), the total mass is \Mpoly.

For the second method, we use the ellipse and width of the stream measured above.  If the stream is circular with a Gaussian profile, then the total light of the stream is equal to the integrated light within one Gaussian width element multiplied by the circumference of the circle.  This is:
\begin{equation}
	f_{\rm total}=S_{\rm max}\sigma(2\pi)^{3/2}\eta R,
\end{equation}
where $f_{\rm total}$ is the total flux density of the stream in Jy, $S_{\rm max}$ is the surface brightness at the peak of the Gaussian fit in Jy~sr$^{-1}$, $\sigma$ is the standard deviation of the Gaussian fit in radians, $\eta$ is the number of times the stream loops, and {\it R} is the radius of the circular orbit in radians.  With the average standard deviation of the Gaussian fit to the stream, and with the semi-major axis of the ellipse as the radius, we find a mass of \Mlight.

The final method used to determine the mass of the stream is the dynamical method of \cite{Johnston+01}.  This method assumes that the stream's orbit is circular and the stream is within a logarithmic potential.  From \cite{Johnston+01}:
\begin{equation}
	M_{\rm total}[M_{\odot}] \sim 10^{11} \left(\frac{w}{R}\right)^3 \left(\frac{R_p}{{\rm 10~kpc}}\right) \left(\frac{v_{\rm circ}}{200~{\rm km~s^{-1}}}\right)^2~M_{\odot},
\end{equation}
where {\it w} is the width of stream at the deprojected radius {\it R}, $R_p$ is the pericentre radius of the stream, and $v_{\rm circ}$ is the circular velocity of the host galaxy.  The width at 20\% of the maximum is \StreamWidth, described above.  The circular velocity is 180~${\rm km~s^{-1}}$ \citep{Bosma78, Battaglia+06}.  With our assumption that the stream's orbit is circular $R_p=R$, and $R=31$~kpc, shown above.  This results in a dynamically derived estimate of the stream's total mass of \Mdyn.

In addition to an estimate of the mass of the stream, we also calculate the time since disruption with the derivation from \cite{Johnston+01}:
\begin{equation}
	t \sim 0.01 \Phi \left(\frac{R}{w}\right) \left(\frac{R_{\rm circ}}{{\rm 10~kpc}}\right) \left(\frac{{\rm 200~km~s^{-1}}}{v_{\rm circ}}\right){\rm Gyr},
\end{equation}
where {\it t} is the time since disruption, $\Phi$ is the angular extent of the stream, and $R_{\rm circ}$ is the circular radius of the stream.  Assuming that the stream is circular with some number of loops, $\Phi=2\pi\eta$. The remaining parameters are described above.  We find a time since disruption of \tdisrup, which agrees with the result of \cite{Chonis+11} of $\sim$1.8$\eta$~Gyr.

\section{Discussion}
\subsection{Inner Stellar Halo}

The surface brightness profile of M63 reveals an extended, low surface brightness feature in the form of an up-bending break beyond $\sim$15~kpc.  These up-bending breaks are not unusual for galaxies of M63's Hubble Type \cite[$T=4$,][]{deVaucouleurs+91}; \cite{Pohlen+Trujillo06} have found that $\sim$50\%\ of local SDSS galaxies with $T=2.5-4.4$ feature up-bending breaks), but what could have caused this break?

Classical surface brightness profiles feature either a single disc component \citep{Patterson40, deVaucouleurs59, Freeman70}, or a truncated, down-bending, profile \citep{vanderKruit79, vanderKruit87}.  These results are reproduced in simulations of disc galaxies in solitary environments.  The simulations work on the principles that gas forms stars after reaching a critical density threshold per the Kennicutt-Schmidt Law \citep{Kennicutt98}, stars may migrate from their original positions through interactions with spiral arms, and that there are no mergers over the galaxy's lifetime.  These assumptions result in discs with down-bending breaks at large radii; due to breaks in the gas-density profile \citep{Roskar+08a, Sanchez-Blazquez+09, Martinez-Serrano+09}, or by stars scattered from interactions with spiral arms \citep{Roskar+09}.  The only time a solitary disc may naturally form an up-bending break is in the presence of a varying galaxy cluster potential \citep{Moore+96, Moore+99}, or in a dark matter halo with low angular momentum \citep{Herpich+15}.  M63 is not within a galaxy cluster, and while there are no measurements of the angular momentum of M63's halo, the simulations that found up-bending breaks from these potentials also have large bulges, which is not seen in M63.  Thus far, simulations of galaxies evolving in solitary environments cannot fully reproduce what we observe with M63.

Simulations and analytical models of galaxy mergers in a $\Lambda$CDM universe find that stellar haloes are produced when satellite galaxies are accreted on to larger galaxies.  The signature of these haloes appear in surface brightness profiles in the form of up-bending breaks at radii of 15-20 kpc \citep{Tissera+14}, and at surface brightnesses dimmer than $28$~AB~mag~arcsec$^{-2}$ \citep{Purcell+07, Cooper+10, Purcell+11, Cooper+13}; the same radial, and surface brightness, regime where we find the up-bending break in M63.  The presence of stellar haloes at these radii and surface brightness is also confirmed observationally for the Milky Way \citep{Carollo+10}, M31 \citep{Courteau+11}, and M81 \citep{Monachesi+13}.

Direct measurements of the observed properties of M63 also suggest that the up-bending break is the signature of a stellar halo.  By stacking many late-type SDSS galaxies into a single mosaic, \cite{D'Souza+14} have detected an average stellar halo.  They find that the $g-r$ colour profile of the stacked galaxies decreases with radius, until it reaches a break point, where the colour reddens sharply.  This is seen in the $B-R$ image of M63 from \cite{Chonis+11}, where the average $B-R$ colour jumps from $\sim$0.8 at the outermost edge of the disc to $\sim$1.2--1.4 beyond the 15~kpc break radius.  This reddening suggests a change from a young stellar population within a disc, to older stellar populations within a halo.  We assume colour is an analog of age at large radii because metallicity gradients have been found to flatten beyond $R_{25}$ \citep{Bresolin13, Bush+14, Kudritzki+14}.  A halo scenario is also supported by the far-ultraviolet map of M63 from \cite{Thilker+07}.  This map finds no far-ultraviolet emission past 15~kpc to a star formation rate density of $3\times10^{-4}M_{\odot}$yr$^{-1}$kpc$^{-2}$, the minimum threshold for star formation \citep{Kennicutt98}.  While these stars may have formed as part of a disc in the past, this is unlikely as simulations have found that the surface density of discs rapidly declines past the star formation threshold \citep{Roskar+09}.  Additionally, there is no detected spiral structure in the $B$, $R$, 3.6~\um, and 4.5~\um\ data, beyond 15~kpc.  The HI maps of \cite{Battaglia+06} contain spiral structure past 15~kpc, however, this structure is inclined in respect to the optical and infrared data in the same radial regime.  We therefore assume that the HI gas is not associated with the optical and infrared emission.  Both simulations and observational evidence points to the break as the signature of a stellar halo, rather than an up-bending break in a disc.  We therefore assume that our halo-model power-law fit is the correct decomposition method for M63.

\begin{figure}
	\epsscale{0.8}
	\plotone{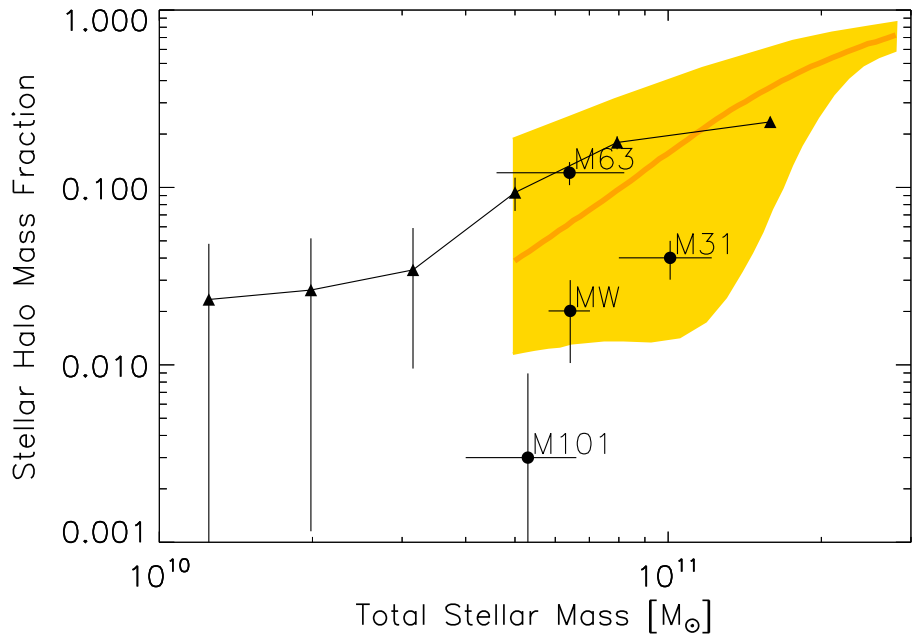}
	\caption{Stellar halo mass fraction versus total stellar mass.  The circles are results for M63 (this work), M31 \citep{Courteau+11}, M101 \citep{vanDokkum+14}, and the Milky Way \citep{Carollo+10}.  The orange line and yellow region are the median and 1~$\sigma$ uncertainty, respectively, from the numerical simulations of \cite{Cooper+13}.  Finally, the triangles are the results from an analysis of SDSS data from \cite{D'Souza+14}.}
	\label{fig:frac}
\end{figure}

With the assumption that the break is due to a stellar halo, we may compare our mass measurement of the halo to simulations based on a $\Lambda$CDM cosmology.  These simulations operate on the base assumption that galaxy evolution is driven by the hierarchical growth of galaxies via the accretion of small satellite galaxies on to larger host galaxies.  A semi-analytical analysis of $\Lambda$CDM based $N$-body simulations \citep{Springel05, Boylan-Kolchin+09} finds that the evidence of this accretion exists within a stellar halo component \cite{Bullock+Johnston05, Purcell+07, Cooper+10, Cooper+13}.  We assume that the bulk of M63's stellar halo mass is due to the accretion of smaller satellite galaxies.  The stellar halo mass fraction derived from stars formed {\it in situ} in the Milky Way is only $\sim$1$\%$ \citep[e.g.][]{Morrison93, Chiba+Beers00, Purcell+07, Bell+08}, whereas the total stellar mass fraction is found to be near $\sim$2$\%$ \citep[e.g.][]{Law+05, Carollo+10}.  If the mechanism for populating the halo with stars born {\it in situ} is similar in the Milky Way and in M63 (they are both spirals of nearly equal mass), then a stellar halo mass fraction above $\sim$1$\%$ is comprised mostly of accreted stars.  In Figure~\ref{fig:frac} we plot our result for M63 along with the results for the Milky Way \citep{Carollo+10}, M31 \citep{Courteau+11}, M81 \citep{Monachesi+13}, and M101 \citep{vanDokkum+14}.  Additionally, we include the results of the analysis of \cite{D'Souza+14}; they measured the average stellar halo mass fraction over many mass bins by stacking thousands of SDSS galaxies.  The model prediction from the numerical simulations of \cite{Cooper+13} is also included.  Our result, and most other results (besides M101), fit well within the model prediction of \cite{Cooper+13} and the results from the analysis of SDSS data \citep{D'Souza+14}. 

We note that populating stellar haloes with stars from accreted satellites is a stochastic process, where most of the mass comes from a few massive dwarf galaxies \citep{Bullock+Johnston05, Cooper+10}.  Thus, a large scatter is to be expected in the stellar halo mass fraction to total stellar mass relation.  It is not surprising to have a result such as M101; this galaxy must not have accreted enough massive dwarfs to produce a halo detectable by \cite{vanDokkum+14}.  We also note that the result from \cite{D'Souza+14} is for the average stellar halo, the uncertainty quoted is based on detecting an average stellar halo, and in no way describes the intrinsic scatter expected from the stochastic accretion of satellite galaxies.  Thus, our results, other observational studies, and model predictions, are consistent with one another within the assumption that the accretion of satellite galaxies is a highly stochastic process.

\subsection{Tidal Stream}

We chose M63 for this study because of the nearby, dramatic, tidal stream.  This feature dominates the spatial extent of the 3.6~\um\ mosaic; the radius of the disc is roughly half that of the stream.  These features are ubiquitous for Milky Way sized spiral galaxies in a $\Lambda$CDM universe, as shown in semi-analytical simulations \citep{Bullock+Johnston05, Purcell+07, Cooper+10, Purcell+11, Cooper+13}.  However, only a handful of galaxies show obvious evidence of streams in the EDGES sample, and only M63 and NGC4013 feature prominent streams.  Yet streams should be common events for galaxies; the remnants of streams, stellar haloes, can be an order of magnitude more massive than an individual stream.  In M63, for example, the halo has \HaloStreamFrac\ times more mass than the stream (assuming the stream loops once).  Given that the mass of the tidal stream is near the upper limit for dwarf galaxies \citep{Mateo+98, Cook+14}, M63 must have accreted many more galaxies than even the halo-to-stream mass fraction suggests.

A simple explanation for the lack of streams is that the accretion rate of satellite galaxies was greater in the past.  To test this explanation, we compare the average accretion rate derived from the stellar halo, to the current accretion rate derived from the stellar stream.  To measure the current accretion rate, we make the following assumptions: all of the mass of the tidal stream's progenitor galaxy will be accreted on to M63, the stream loops a single time, and over the timescale since disruption the tidal stream is the only contributor to the accretion rate.  Thus, the current accretion rate is the mass of the progenitor divided by the time since disruption, see \S~\ref{sec:tidal_stream} for these measurements.  This is an upper limit because the tidal stream will be contributing to the mass of the halo past the time since disruption.  To find a lower limit for the average accretion rate, we assume that the halo is comprised entirely of accreted stars, and the age of M63 is the age of the Universe, $\sim$13.5~Gyr.  Thus, the average accretion rate is the mass of the stellar halo divided by 13.5~Gyr, see \S~\ref{sec:prof_fit} for this measurement.  The ratio of the average accretion rate to the current accretion rate is at least \RateFrac.  Given the many assumptions of this analysis, the rate at which M63 is accreting matter from this dramatic tidal event is much lower than in the past.  This is in agreement with simulations, which find that the majority of the mass which forms the stellar halo was accreted in the first $\sim$5~Gyr of a galaxy's existence \citep{Bullock+Johnston05, Cooper+10}.

\section{Conclusion}

We present an analysis of the low surface brightness NIR properties of the flocculent spiral M63 as part of the EDGES Survey.  We use the 3.6~\um\ data to derive the mass of the old stellar population within the individual components of the galaxy including the bulge, disc, halo and nearby tidal stream.

The M63 mosaic consists of data from the EDGES Survey, an ultra-deep (1800~s~pixel$^{-1}$), wide-field ($>5~R_{25}$), Spitzer warm mission survey of 92 nearby galaxies \citep{vanZee+11}.  The resulting mosaic reaches a depth of 29~AB~mag~arcsec$^{-2}$ at a distance of $\sim3~R_{25}$ on the surface brightness profile. At this imaging depth the outer component of the galaxy is detected (an outer disc or inner stellar halo) in the form of an up-bending break in the surface brightness profile, along with the nearby tidal stream.

Several factors indicate that the outer component is a stellar halo.  An up-bending break is unlikely for a disc as M63 is not within a galaxy cluster, the colour of the feature suggests an old stellar population, there is little active star formation in this region, and there is no evidence of spiral structure past 15~kpc.  The halo to total mass ratio is \Mfrac, the largest halo mass ratio recorded thus far for an individual galaxy.  This ratio agrees well with the ratios derived from SDSS data for the average galaxy \citep{D'Souza+14}, and is within the envelope predicted by the semi-analytical methods of \citep{Cooper+13}.  With the EDGES survey of 92 galaxies we will be able to construct a large sample of stellar halo measurements for a more statistically-grounded result \citep[in prep.]{Staudaher+15}.  These new observations will shed light on the interplay between large galaxies and their satellites, and will have interesting implications for the missing satellite problem.

The nearby tidal stream has a mass of \Mlight, derived from the stream's luminosity and width, and \Mdyn, measured dynamically.  The dynamical method also finds a time since disruption of \tdisrup\ for this stream.  Despite the prominence of the tidal stream, the \HaloStreamFrac\ times more massive halo suggests that the accretion rate of satellites was much larger in the past.  This is supported by the ratio of the current to past accretion rates; the average accretion rate is at least \RateFrac\ times the accretion rate derived from the stream alone.  This is not surprising given that the Universe was much denser, and galaxies were smaller in the past.  Semi-analytical models of Milky Way analogs also find that satellite galaxies were, in general, accreted more rapidly in the past \citep{Bullock+Johnston05, Cooper+10}.

This work is based on observations made with the Spitzer Space Telescope, which is operated by the Jet Propulsion Laboratory, California Institute of Technology Support under a contract with NASA. Support for this work was provided by NASA through an award issued by the JPL/Caltech. This research has made use of the NASA/IPAC Infrared Science Archive, which is operated by the Jet Propulsion Laboratory, California Institute of Technology, under contract with NASA. This research has made use of the NASA/IPAC Extragalactic Database (NED) which is operated by the Jet Propulsion Laboratory, California Institute of Technology, under contract with the National Aeronautics and Space Administration.  We would also like to thank the anonymous referee for their constructive feedback.
 
\clearpage
\bibliographystyle{apj}
\bibliography{refs}

\newcommand{\noop}[1]{}
\begin{thebibliography}{}
\expandafter\ifx\csname natexlab\endcsname\relax\def\natexlab#1{#1}\fi

\bibitem[{{Abraham} \& {van Dokkum}(2014)}]{Abraham+vanDokkum14}
{Abraham}, R.~G., \& {van Dokkum}, P.~G. 2014, \pasp, 126, 55

\bibitem[{{Bailin} {et~al.}(2011){Bailin}, {Bell}, {Chappell}, {Radburn-Smith},
  \& {de Jong}}]{Bailin+11}
{Bailin}, J., {Bell}, E.~F., {Chappell}, S.~N., {Radburn-Smith}, D.~J., \& {de
  Jong}, R.~S. 2011, \apj, 736, 24

\bibitem[{{Barnes} {et~al.}(2014){Barnes}, {van Zee}, {Dale}, {Staudaher},
  {Bullock}, {Calzetti}, {Chandar}, \& {Dalcanton}}]{Barnes+14}
{Barnes}, K.~L., {van Zee}, L., {Dale}, D.~A., {et~al.} 2014, \apj, 789, 126

\bibitem[{{Battaglia} {et~al.}(2006){Battaglia}, {Fraternali}, {Oosterloo}, \&
  {Sancisi}}]{Battaglia+06}
{Battaglia}, G., {Fraternali}, F., {Oosterloo}, T., \& {Sancisi}, R. 2006,
  \aap, 447, 49

\bibitem[{{Bell} \& {de Jong}(2001)}]{Bell+deJong01}
{Bell}, E.~F., \& {de Jong}, R.~S. 2001, \apj, 550, 212

\bibitem[{{Bell} {et~al.}(2008){Bell}, {Zucker}, {Belokurov}, {Sharma},
  {Johnston}, {Bullock}, {Hogg}, {Jahnke}, {de Jong}, {Beers}, {Evans},
  {Grebel}, {Ivezi{\'c}}, {Koposov}, {Rix}, {Schneider}, {Steinmetz}, \&
  {Zolotov}}]{Bell+08}
{Bell}, E.~F., {Zucker}, D.~B., {Belokurov}, V., {et~al.} 2008, \apj, 680, 295

\bibitem[{{Belokurov} {et~al.}(2006){Belokurov}, {Zucker}, {Evans}, {Gilmore},
  {Vidrih}, {Bramich}, {Newberg}, {Wyse}, {Irwin}, {Fellhauer}, {Hewett},
  {Walton}, {Wilkinson}, {Cole}, {Yanny}, {Rockosi}, {Beers}, {Bell},
  {Brinkmann}, {Ivezi{\'c}}, \& {Lupton}}]{Belokurov+06}
{Belokurov}, V., {Zucker}, D.~B., {Evans}, N.~W., {et~al.} 2006, \apjl, 642,
  L137

\bibitem[{{Borlaff} {et~al.}(2014){Borlaff}, {Eliche-Moral},
  {Rodr{\'{\i}}guez-P{\'e}rez}, {Querejeta}, {Tapia}, {P{\'e}rez-Gonz{\'a}lez},
  {Zamorano}, {Gallego}, \& {Beckman}}]{Borlaff+14}
{Borlaff}, A., {Eliche-Moral}, M.~C., {Rodr{\'{\i}}guez-P{\'e}rez}, C.,
  {et~al.} 2014, \aap, 570, A103

\bibitem[{{Bosma}(1978)}]{Bosma78}
{Bosma}, A. 1978, PhD thesis, PhD Thesis, Groningen Univ., (1978)

\bibitem[{{Boylan-Kolchin} {et~al.}(2009){Boylan-Kolchin}, {Springel}, {White},
  {Jenkins}, \& {Lemson}}]{Boylan-Kolchin+09}
{Boylan-Kolchin}, M., {Springel}, V., {White}, S.~D.~M., {Jenkins}, A., \&
  {Lemson}, G. 2009, \mnras, 398, 1150

\bibitem[{{Bresolin}(2013)}]{Bresolin13}
{Bresolin}, F. 2013, \apjl, 772, L23

\bibitem[{{Bullock} \& {Johnston}(2005)}]{Bullock+Johnston05}
{Bullock}, J.~S., \& {Johnston}, K.~V. 2005, \apj, 635, 931

\bibitem[{{Bush} {et~al.}(2014){Bush}, {Kennicutt}, {Ashby}, {Johnson},
  {Bresolin}, \& {Fazio}}]{Bush+14}
{Bush}, S.~J., {Kennicutt}, R.~C., {Ashby}, M.~L.~N., {et~al.} 2014, \apj, 793,
  65

\bibitem[{{Carollo} {et~al.}(2010){Carollo}, {Beers}, {Chiba}, {Norris},
  {Freeman}, {Lee}, {Ivezi{\'c}}, {Rockosi}, \& {Yanny}}]{Carollo+10}
{Carollo}, D., {Beers}, T.~C., {Chiba}, M., {et~al.} 2010, \apj, 712, 692

\bibitem[{{Chiba} \& {Beers}(2000)}]{Chiba+Beers00}
{Chiba}, M., \& {Beers}, T.~C. 2000, \aj, 119, 2843

\bibitem[{{Chilingarian} {et~al.}(2010){Chilingarian}, {Di Matteo}, {Combes},
  {Melchior}, \& {Semelin}}]{Chilingarian+10}
{Chilingarian}, I.~V., {Di Matteo}, P., {Combes}, F., {Melchior}, A.-L., \&
  {Semelin}, B. 2010, \aap, 518, A61

\bibitem[{{Chonis} {et~al.}(2011){Chonis}, {Mart{\'{\i}}nez-Delgado}, {Gabany},
  {Majewski}, {Hill}, {Gralak}, \& {Trujillo}}]{Chonis+11}
{Chonis}, T.~S., {Mart{\'{\i}}nez-Delgado}, D., {Gabany}, R.~J., {et~al.} 2011,
  \aj, 142, 166

\bibitem[{{Cook} {et~al.}(2014){Cook}, {Dale}, {Johnson}, {Van Zee}, {Lee},
  {Kennicutt}, {Calzetti}, {Staudaher}, \& {Engelbracht}}]{Cook+14}
{Cook}, D.~O., {Dale}, D.~A., {Johnson}, B.~D., {et~al.} 2014, \mnras, 445, 899

\bibitem[{{Cooper} {et~al.}(2013){Cooper}, {D'Souza}, {Kauffmann}, {Wang},
  {Boylan-Kolchin}, {Guo}, {Frenk}, \& {White}}]{Cooper+13}
{Cooper}, A.~P., {D'Souza}, R., {Kauffmann}, G., {et~al.} 2013, \mnras, 434,
  3348

\bibitem[{{Cooper} {et~al.}(2010){Cooper}, {Cole}, {Frenk}, {White}, {Helly},
  {Benson}, {De Lucia}, {Helmi}, {Jenkins}, {Navarro}, {Springel}, \&
  {Wang}}]{Cooper+10}
{Cooper}, A.~P., {Cole}, S., {Frenk}, C.~S., {et~al.} 2010, \mnras, 406, 744

\bibitem[{{Courteau} {et~al.}(2011){Courteau}, {Widrow}, {McDonald},
  {Guhathakurta}, {Gilbert}, {Zhu}, {Beaton}, \& {Majewski}}]{Courteau+11}
{Courteau}, S., {Widrow}, L.~M., {McDonald}, M., {et~al.} 2011, \apj, 739, 20

\bibitem[{{Dale} {et~al.}(2009){Dale}, {Cohen}, {Johnson}, {Schuster},
  {Calzetti}, {Engelbracht}, {Gil de Paz}, {Kennicutt}, {Lee}, {Begum},
  {Block}, {Dalcanton}, {Funes}, {Gordon}, {Johnson}, {Marble}, {Sakai},
  {Skillman}, {van Zee}, {Walter}, {Weisz}, {Williams}, {Wu}, \&
  {Wu}}]{Dale+09}
{Dale}, D.~A., {Cohen}, S.~A., {Johnson}, L.~C., {et~al.} 2009, \apj, 703, 517

\bibitem[{{de Vaucouleurs}(1959)}]{deVaucouleurs59}
{de Vaucouleurs}, G. 1959, Handbuch der Physik, 53, 311

\bibitem[{{de Vaucouleurs} {et~al.}(1991){de Vaucouleurs}, {de Vaucouleurs},
  {Corwin}, {Buta}, {Paturel}, \& {Fouqu{\'e}}}]{deVaucouleurs+91}
{de Vaucouleurs}, G., {de Vaucouleurs}, A., {Corwin}, Jr., H.~G., {et~al.}
  1991, {Third Reference Catalogue of Bright Galaxies. Volume I: Explanations
  and references. Volume II: Data for galaxies between 0$^{h}$ and 12$^{h}$.
  Volume III: Data for galaxies between 12$^{h}$ and 24$^{h}$.}

\bibitem[{{D'Souza} {et~al.}(2014){D'Souza}, {Kauffman}, {Wang}, \&
  {Vegetti}}]{D'Souza+14}
{D'Souza}, R., {Kauffman}, G., {Wang}, J., \& {Vegetti}, S. 2014, \mnras, 443,
  1433

\bibitem[{{Eskew} {et~al.}(2012){Eskew}, {Zaritsky}, \& {Meidt}}]{Eskew+12}
{Eskew}, M., {Zaritsky}, D., \& {Meidt}, S. 2012, \aj, 143, 139

\bibitem[{{Freeman}(1970)}]{Freeman70}
{Freeman}, K.~C. 1970, \apj, 160, 811

\bibitem[{{Graham} \& {Driver}(2005)}]{Graham+Driver05}
{Graham}, A.~W., \& {Driver}, S.~P. 2005, \pasa, 22, 118

\bibitem[{{Gratton} {et~al.}(2012){Gratton}, {Carretta}, \&
  {Bragaglia}}]{Gratton+12}
{Gratton}, R.~G., {Carretta}, E., \& {Bragaglia}, A. 2012, \aapr, 20, 50

\bibitem[{{Herpich} {et~al.}(2015){Herpich}, {Stinson}, {Dutton}, {Rix},
  {Martig}, {Ro{\v s}kar}, {Macci{\`o}}, {Quinn}, \& {Wadsley}}]{Herpich+15}
{Herpich}, J., {Stinson}, G.~S., {Dutton}, A.~A., {et~al.} 2015, ArXiv
  e-prints, arXiv:1501.01960

\bibitem[{{Hopkins} {et~al.}(2006){Hopkins}, {Hernquist}, {Cox}, {Di Matteo},
  {Robertson}, \& {Springel}}]{Hopkins+06}
{Hopkins}, P.~F., {Hernquist}, L., {Cox}, T.~J., {et~al.} 2006, \apjs, 163, 1

\bibitem[{{Ibata} {et~al.}(1994){Ibata}, {Gilmore}, \& {Irwin}}]{Ibata+94}
{Ibata}, R.~A., {Gilmore}, G., \& {Irwin}, M.~J. 1994, \nat, 370, 194

\bibitem[{{Johnson} \& {Pilachowski}(2010)}]{Johnson+10}
{Johnson}, C.~I., \& {Pilachowski}, C.~A. 2010, \apj, 722, 1373

\bibitem[{{Johnston} {et~al.}(2001){Johnston}, {Sackett}, \&
  {Bullock}}]{Johnston+01}
{Johnston}, K.~V., {Sackett}, P.~D., \& {Bullock}, J.~S. 2001, \apj, 557, 137

\bibitem[{{Kennicutt}(1998)}]{Kennicutt98}
{Kennicutt}, Jr., R.~C. 1998, \araa, 36, 189

\bibitem[{{Krick} {et~al.}(2011){Krick}, {Bridge}, {Desai}, {Mihos}, {Murphy},
  {Rudick}, {Surace}, \& {Neill}}]{Krick+11}
{Krick}, J.~E., {Bridge}, C., {Desai}, V., {et~al.} 2011, \apj, 735, 76

\bibitem[{{Kudritzki} {et~al.}(2014){Kudritzki}, {Urbaneja}, {Bresolin},
  {Hosek}, \& {Przybilla}}]{Kudritzki+14}
{Kudritzki}, R.-P., {Urbaneja}, M.~A., {Bresolin}, F., {Hosek}, Jr., M.~W., \&
  {Przybilla}, N. 2014, \apj, 788, 56

\bibitem[{{Law} {et~al.}(2005){Law}, {Johnston}, \& {Majewski}}]{Law+05}
{Law}, D.~R., {Johnston}, K.~V., \& {Majewski}, S.~R. 2005, \apj, 619, 807

\bibitem[{{Mackey} {et~al.}(2010){Mackey}, {Huxor}, {Ferguson}, {Irwin},
  {Tanvir}, {McConnachie}, {Ibata}, {Chapman}, \& {Lewis}}]{Mackey+10}
{Mackey}, A.~D., {Huxor}, A.~P., {Ferguson}, A.~M.~N., {et~al.} 2010, \apjl,
  717, L11

\bibitem[{{Marino} {et~al.}(2010){Marino}, {Piotto}, {Gratton}, {Milone},
  {Zoccali}, {Bedin}, {Villanova}, \& {Bellini}}]{Marino+10}
{Marino}, A.~F., {Piotto}, G., {Gratton}, R., {et~al.} 2010, in IAU Symposium,
  Vol. 268, IAU Symposium, ed. C.~{Charbonnel}, M.~{Tosi}, F.~{Primas}, \&
  C.~{Chiappini}, 183--184

\bibitem[{{Markwardt}(2009)}]{Markwardt09}
{Markwardt}, C.~B. 2009, in Astronomical Society of the Pacific Conference
  Series, Vol. 411, Astronomical Data Analysis Software and Systems XVIII, ed.
  D.~A. {Bohlender}, D.~{Durand}, \& P.~{Dowler}, 251

\bibitem[{{Martin} {et~al.}(2014){Martin}, {Ibata}, {Rich}, {Collins},
  {Fardal}, {Irwin}, {Lewis}, {McConnachie}, {Babul}, {Bate}, {Chapman},
  {Conn}, {Crnojevi{\'c}}, {Ferguson}, {Mackey}, {Navarro}, {Pe{\~n}arrubia},
  {Tanvir}, \& {Valls-Gabaud}}]{Martin+14}
{Martin}, N.~F., {Ibata}, R.~A., {Rich}, R.~M., {et~al.} 2014, \apj, 787, 19

\bibitem[{{Mart{\'{\i}}nez-Delgado} {et~al.}(2014){Mart{\'{\i}}nez-Delgado},
  {D'Onghia}, {Chonis}, {Beaton}, {Teuwen}, {GaBany}, {Grebel}, \&
  {Morales}}]{Martinez-Delgado+14}
{Mart{\'{\i}}nez-Delgado}, D., {D'Onghia}, E., {Chonis}, T.~S., {et~al.} 2014,
  ArXiv e-prints, arXiv:1410.6368

\bibitem[{{Mart{\'{\i}}nez-Delgado} {et~al.}(2008){Mart{\'{\i}}nez-Delgado},
  {Pe{\~n}arrubia}, {Gabany}, {Trujillo}, {Majewski}, \&
  {Pohlen}}]{Martinez-Delgado+08}
{Mart{\'{\i}}nez-Delgado}, D., {Pe{\~n}arrubia}, J., {Gabany}, R.~J., {et~al.}
  2008, \apj, 689, 184

\bibitem[{{Mart{\'{\i}}nez-Delgado} {et~al.}(2009){Mart{\'{\i}}nez-Delgado},
  {Pohlen}, {Gabany}, {Majewski}, {Pe{\~n}arrubia}, \&
  {Palma}}]{Martinez-Delgado+09}
{Mart{\'{\i}}nez-Delgado}, D., {Pohlen}, M., {Gabany}, R.~J., {et~al.} 2009,
  \apj, 692, 955

\bibitem[{{Mart{\'{\i}}nez-Delgado} {et~al.}(2010){Mart{\'{\i}}nez-Delgado},
  {Gabany}, {Crawford}, {Zibetti}, {Majewski}, {Rix}, {Fliri},
  {Carballo-Bello}, {Bardalez-Gagliuffi}, {Pe{\~n}arrubia}, {Chonis}, {Madore},
  {Trujillo}, {Schirmer}, \& {McDavid}}]{Martinez-Delgado+10}
{Mart{\'{\i}}nez-Delgado}, D., {Gabany}, R.~J., {Crawford}, K., {et~al.} 2010,
  \aj, 140, 962

\bibitem[{{Mart{\'{\i}}nez-Delgado} {et~al.}(2012){Mart{\'{\i}}nez-Delgado},
  {Romanowsky}, {Gabany}, {Annibali}, {Arnold}, {Fliri}, {Zibetti}, {van der
  Marel}, {Rix}, {Chonis}, {Carballo-Bello}, {Aloisi}, {Macci{\`o}},
  {Gallego-Laborda}, {Brodie}, \& {Merrifield}}]{Martinez-Delgado+12}
{Mart{\'{\i}}nez-Delgado}, D., {Romanowsky}, A.~J., {Gabany}, R.~J., {et~al.}
  2012, \apjl, 748, L24

\bibitem[{{Mart{\'{\i}}nez-Serrano} {et~al.}(2009){Mart{\'{\i}}nez-Serrano},
  {Serna}, {Dom{\'e}nech-Moral}, \&
  {Dom{\'{\i}}nguez-Tenreiro}}]{Martinez-Serrano+09}
{Mart{\'{\i}}nez-Serrano}, F.~J., {Serna}, A., {Dom{\'e}nech-Moral}, M., \&
  {Dom{\'{\i}}nguez-Tenreiro}, R. 2009, \apjl, 705, L133

\bibitem[{{Mateo}(1998)}]{Mateo+98}
{Mateo}, M.~L. 1998, \araa, 36, 435

\bibitem[{{Mathewson} {et~al.}(1974){Mathewson}, {Cleary}, \&
  {Murray}}]{Mathewson+74}
{Mathewson}, D.~S., {Cleary}, M.~N., \& {Murray}, J.~D. 1974, \apj, 190, 291

\bibitem[{{Meidt} {et~al.}(2014){Meidt}, {Schinnerer}, {van de Ven},
  {Zaritsky}, {Peletier}, {Knapen}, {Sheth}, {Regan}, {Querejeta},
  {Mu{\~n}oz-Mateos}, {Kim}, {Hinz}, {Gil de Paz}, {Athanassoula}, {Bosma},
  {Buta}, {Cisternas}, {Ho}, {Holwerda}, {Skibba}, {Laurikainen}, {Salo},
  {Gadotti}, {Laine}, {Erroz-Ferrer}, {Comer{\'o}n}, {Men{\'e}ndez-Delmestre},
  {Seibert}, \& {Mizusawa}}]{Meidt+14}
{Meidt}, S.~E., {Schinnerer}, E., {van de Ven}, G., {et~al.} 2014, \apj, 788,
  144

\bibitem[{{Milone} {et~al.}(2010){Milone}, {Piotto}, {King}, {Bedin},
  {Anderson}, {Marino}, {Momany}, {Malavolta}, \& {Villanova}}]{Milone+10}
{Milone}, A.~P., {Piotto}, G., {King}, I.~R., {et~al.} 2010, \apj, 709, 1183

\bibitem[{{Monachesi} {et~al.}(2013){Monachesi}, {Bell}, {Radburn-Smith},
  {Vlaji{\'c}}, {de Jong}, {Bailin}, {Dalcanton}, {Holwerda}, \&
  {Streich}}]{Monachesi+13}
{Monachesi}, A., {Bell}, E.~F., {Radburn-Smith}, D.~J., {et~al.} 2013, \apj,
  766, 106

\bibitem[{{Moore} {et~al.}(1996){Moore}, {Katz}, {Lake}, {Dressler}, \&
  {Oemler}}]{Moore+96}
{Moore}, B., {Katz}, N., {Lake}, G., {Dressler}, A., \& {Oemler}, A. 1996,
  \nat, 379, 613

\bibitem[{{Moore} {et~al.}(1999){Moore}, {Lake}, {Quinn}, \&
  {Stadel}}]{Moore+99}
{Moore}, B., {Lake}, G., {Quinn}, T., \& {Stadel}, J. 1999, \mnras, 304, 465

\bibitem[{{Morrison}(1993)}]{Morrison93}
{Morrison}, H.~L. 1993, \aj, 106, 578

\bibitem[{{Nasonova} {et~al.}(2011){Nasonova}, {de Freitas Pacheco}, \&
  {Karachentsev}}]{Nasonova+11}
{Nasonova}, O.~G., {de Freitas Pacheco}, J.~A., \& {Karachentsev}, I.~D. 2011,
  \aap, 532, A104

\bibitem[{{Oh} {et~al.}(2008){Oh}, {de Blok}, {Walter}, {Brinks}, \&
  {Kennicutt}}]{Oh+08}
{Oh}, S.-H., {de Blok}, W.~J.~G., {Walter}, F., {Brinks}, E., \& {Kennicutt},
  Jr., R.~C. 2008, \aj, 136, 2761

\bibitem[{{Patterson}(1940)}]{Patterson40}
{Patterson}, F.~S. 1940, Harvard College Observatory Bulletin, 914, 9

\bibitem[{{Pohlen} \& {Trujillo}(2006)}]{Pohlen+Trujillo06}
{Pohlen}, M., \& {Trujillo}, I. 2006, \aap, 454, 759

\bibitem[{{Purcell} {et~al.}(2011){Purcell}, {Bullock}, {Tollerud}, {Rocha}, \&
  {Chakrabarti}}]{Purcell+11}
{Purcell}, C.~W., {Bullock}, J.~S., {Tollerud}, E.~J., {Rocha}, M., \&
  {Chakrabarti}, S. 2011, \nat, 477, 301

\bibitem[{{Purcell} {et~al.}(2007){Purcell}, {Bullock}, \&
  {Zentner}}]{Purcell+07}
{Purcell}, C.~W., {Bullock}, J.~S., \& {Zentner}, A.~R. 2007, \apj, 666, 20

\bibitem[{{Radburn-Smith} {et~al.}(2011){Radburn-Smith}, {de Jong}, {Seth},
  {Bailin}, {Bell}, {Brown}, {Bullock}, {Courteau}, {Dalcanton}, {Ferguson},
  {Goudfrooij}, {Holfeltz}, {Holwerda}, {Purcell}, {Sick}, {Streich}, {Vlajic},
  \& {Zucker}}]{Radburn-Smith+11}
{Radburn-Smith}, D.~J., {de Jong}, R.~S., {Seth}, A.~C., {et~al.} 2011, \apjs,
  195, 18

\bibitem[{{Reach} {et~al.}(2005){Reach}, {Megeath}, {Cohen}, {Hora}, {Carey},
  {Surace}, {Willner}, {Barmby}, {Wilson}, {Glaccum}, {Lowrance}, {Marengo}, \&
  {Fazio}}]{Reach+05}
{Reach}, W.~T., {Megeath}, S.~T., {Cohen}, M., {et~al.} 2005, \pasp, 117, 978

\bibitem[{{Ro{\v s}kar} {et~al.}(2009){Ro{\v s}kar}, {Debattista}, {Quinn},
  {Stinson}, {Wadsley}, \& {Kaufmann}}]{Roskar+09}
{Ro{\v s}kar}, R., {Debattista}, V.~P., {Quinn}, T.~R., {et~al.} 2009, in IAU
  Symposium, Vol. 254, IAU Symposium, ed. J.~{Andersen}, {Nordstr{\"o}ara},
  B.~{m}, \& J.~{Bland-Hawthorn}, 64P

\bibitem[{{Ro{\v s}kar} {et~al.}(2008){Ro{\v s}kar}, {Debattista}, {Stinson},
  {Quinn}, {Kaufmann}, \& {Wadsley}}]{Roskar+08a}
{Ro{\v s}kar}, R., {Debattista}, V.~P., {Stinson}, G.~S., {et~al.} 2008, \apjl,
  675, L65

\bibitem[{{S{\'a}nchez-Bl{\'a}zquez} {et~al.}(2009){S{\'a}nchez-Bl{\'a}zquez},
  {Courty}, {Gibson}, \& {Brook}}]{Sanchez-Blazquez+09}
{S{\'a}nchez-Bl{\'a}zquez}, P., {Courty}, S., {Gibson}, B.~K., \& {Brook},
  C.~B. 2009, \mnras, 398, 591

\bibitem[{{S{\'e}rsic}(1963)}]{Sersic63}
{S{\'e}rsic}, J.~L. 1963, Boletin de la Asociacion Argentina de Astronomia La
  Plata Argentina, 6, 41

\bibitem[{{S{\'e}rsic}(1968)}]{Sersic68}
---. 1968, {Atlas de galaxias australes}

\bibitem[{{Sonnenfeld} {et~al.}(2014){Sonnenfeld}, {Nipoti}, \&
  {Treu}}]{Sonnenfeld+14}
{Sonnenfeld}, A., {Nipoti}, C., \& {Treu}, T. 2014, \apj, 786, 89

\bibitem[{{Springel}(2005)}]{Springel05}
{Springel}, V. 2005, \mnras, 364, 1105

\bibitem[{{Staudaher}(2015)}]{Staudaher+15}
{Staudaher}, S. 2015, \mnras

\bibitem[{{Thilker} {et~al.}(2007){Thilker}, {Bianchi}, {Meurer}, {Gil de Paz},
  {Boissier}, {Madore}, {Boselli}, {Ferguson}, {Mu{\~n}oz-Mateos}, {Madsen},
  {Hameed}, {Overzier}, {Forster}, {Friedman}, {Martin}, {Morrissey}, {Neff},
  {Schiminovich}, {Seibert}, {Small}, {Wyder}, {Donas}, {Heckman}, {Lee},
  {Milliard}, {Rich}, {Szalay}, {Welsh}, \& {Yi}}]{Thilker+07}
{Thilker}, D.~A., {Bianchi}, L., {Meurer}, G., {et~al.} 2007, \apjs, 173, 538

\bibitem[{{Tissera} {et~al.}(2014){Tissera}, {Beers}, {Carollo}, \&
  {Scannapieco}}]{Tissera+14}
{Tissera}, P.~B., {Beers}, T.~C., {Carollo}, D., \& {Scannapieco}, C. 2014,
  \mnras, 439, 3128

\bibitem[{{Tully} {et~al.}(2009){Tully}, {Rizzi}, {Shaya}, {Courtois},
  {Makarov}, \& {Jacobs}}]{Tully+09}
{Tully}, R.~B., {Rizzi}, L., {Shaya}, E.~J., {et~al.} 2009, \aj, 138, 323

\bibitem[{{van der Kruit}(1979)}]{vanderKruit79}
{van der Kruit}, P.~C. 1979, \aaps, 38, 15

\bibitem[{{van der Kruit}(1987)}]{vanderKruit87}
---. 1987, \aap, 173, 59

\bibitem[{{van Dokkum} {et~al.}(2014){van Dokkum}, {Abraham}, \&
  {Merritt}}]{vanDokkum+14}
{van Dokkum}, P.~G., {Abraham}, R., \& {Merritt}, A. 2014, \apjl, 782, L24

\bibitem[{{van Zee} {et~al.}(2011){van Zee}, {Dale}, {Barnes}, {Staudaher},
  {Calzetti}, {Dalcanton}, {Bullock}, \& {Chandar}}]{vanZee+11}
{van Zee}, L., {Dale}, D.~A., {Barnes}, K.~L., {et~al.} 2011, {Stellar
  Distributions in Dark Matter Halos: Looking Over the Edge}, spitzer Proposal

\bibitem[{{van Zee} {et~al.}(2009){van Zee}, {Dale}, {Barnes}, {Bullock},
  {Calzetti}, {Chandar}, {Dalcanton}, {Dale}, \& {Hinz}}]{vanZee+09}
---. 2009, {Faint Stellar Distributions in Extended HI Disks}, spitzer Proposal

\bibitem[{{Wannier} \& {Wrixon}(1972)}]{Wannier+72}
{Wannier}, P., \& {Wrixon}, G.~T. 1972, \apjl, 173, L119

\end{thebibliography}

\begin{deluxetable}{cccc}
\tablecolumns{4}
\tablewidth{0pc}
\tablecaption{Surface Brightness Fit Parameters \label{table:fit}}
\tablehead{
  \colhead{Component} & \colhead{$I$ [MJy~sr$^{-1}$]} & \colhead{$h$ [$\arcmin$]} & \colhead{Mass [$M_{\odot}$]}}
\startdata
  Bulge &       2.90 $\pm$       0.30 &      0.296 $\pm$      0.026 &    8.0 $\pm$    3.5 $\times$ 10$^9$ \\
  Disc$_1$ &       1.08 $\pm$       0.10 &       1.30 $\pm$       0.10 &  \hspace{1 pt}  3.0 $\pm$    1.3 $\times$ 10$^{10}$ \\
  Disc$_2$ &      0.129 $\pm$      0.031 &       3.48 $\pm$       0.22 &  \hspace{1 pt}  2.6 $\pm$    1.2 $\times$ 10$^{10}$ \\ \hline
  Bulge &       2.65 $\pm$       0.29 &       0.37 $\pm$       0.16 &  \hspace{1 pt} 1.15 $\pm$    0.49 $\times$ 10$^{10}$ \\
  Disc &      0.744 $\pm$      0.048 &      1.87 $\pm$      0.05 & \hspace{1 pt}   4.3 $\pm$    1.7 $\times$ 10$^{10}$ \\
  Halo &    0.00147 $\pm$    0.00016 &  \nodata &    7.7 $\pm$  3.1 $\times$ 10$^9$ \\
\enddata
\tablecomments{This table contains the parameters for each component of the fits to the surface brightness profiles (see \S\ref{sec:prof_fit}). The data above the line designates the model with a S\'ersic bulge with two discs, and the data below the line designates the model with a S\'ersic bulge, disc, and a power-law halo.  $I$ is the central surface brightness of the fit.  $h$ is the scale length of the fit.  The mass is the mass of the component, as discussed in \S\ref{sec:prof_fit}.}
\end{deluxetable}

\end{document}